**Pick-and-place transfer of arbitrary-metal electrodes for van der Waals device fabrication**

Kaijian Xing[1,†,*], Daniel McEwen[1,2,†], Weiyao Zhao[2,3], Abdulhakim Bake[2,4], David Cortie[2,5,6], Jingying Liu[7,8], Thi-Hai-Yen Vu[1,2], James Hone[9], Alastair Stacey[10,11], Mark T. Edmonds[1,2], Kenji Watanabe[12], Takashi Taniguchi[13], Qingdong Ou[3,7,8,*], Dong-Chen Qi[14,15,*], Michael S. Fuhrer[1,2,*]

[1] School of Physics and Astronomy, Monash University, Clayton, Victoria 3800, Australia

[2] Australian Research Council Centre of Excellence in Future Low-Energy Electronics Technologies (FLEET), Monash University, Clayton, Victoria 3800, Australia

[3] Department of Materials Science & Engineering, Monash University, Clayton, Victoria 3800, Australia

[4] Institute for Superconducting and Electric Materials (ISEM), University of Wollongong, Wollongong, NSW, 2522, Australia

[5] School of Physics, University of Wollongong, NSW, 2522, Australia

[6] The Australia Nuclear Science and Technology Organisation (ANSTO), Lucas Heights, NSW, 2234, Australia

[7] Macau University of Science and Technology Zhuhai MUST Science and Technology Research Institute, Zhuhai, 519099, China

[8] Macao Institute of Materials Science and Engineering (MIMSE), Faculty of Innovation Engineering, Macau University of Science and Technology, Taipa, Macao, 999078, China

[9] Department of Mechanical Engineering, Columbia University, New York, New York 10027, United States

[10] School of Science, RMIT University, Melbourne, Victoria 3000, Australia

[11] Princeton Plasma Physics Laboratory, 100 Stellarator Road, Princeton, New Jersey 08540, USA

[12] Research Center for Electronic and Optical Materials, National Institute for Materials Science, 1-1 Namiki, Tsukuba 305-0044, Japan

[13] Research Center for Materials Nanoarchitectonics, National Institute for Materials Science, 1-1 Namiki, Tsukuba 305-0044, Japan

[14] Centre for Materials Science, Queensland University of Technology, Brisbane, Queensland 4001, Australia

[15] School of Chemistry and Physics, Queensland University of Technology, Brisbane, Queensland 4001, Australia

† these two authors contribute equally to this work

* michael.fuhrer@monash.edu;

  dongchen.qi@qut.edu.au;

  qdou@must.edu.mo;

  kaijian.xing@monash.edu

**ABSTRACT:**

Van der Waals electrode integration is a promising strategy to create near-perfect interfaces between metals and two-dimensional materials, with advantages such as eliminating Fermi-level pinning and reducing contact resistance. However, the lack of a simple, generalizable pick-and-place transfer technology has greatly hampered the wide use of this technique. We demonstrate the pick-and-place transfer of pre-fabricated electrodes from reusable polished hydrogenated diamond substrates without the use of any surface treatments or sacrificial layers. The technique enables transfer of large-scale arbitrary metal electrodes, as demonstrated by successful transfer of eight different elemental metals with work functions ranging from 4.22 to 5.65 eV. The mechanical transfer of metal electrodes from diamond onto van der Waals materials creates atomically smooth interfaces with no interstitial impurities or disorder, as observed with cross-sectional high-resolution transmission electron microscopy and energy-dispersive X-ray spectroscopy. As a demonstration of its device application, we use the diamond-transfer technique to create metal contacts to monolayer transition metal dichalcogenide semiconductors with high-work-function Pd, low-work-function Ti, and semi metal Bi to create *n*- and *p*-type field-effect transistors with low Schottky barrier heights. We also extend this technology to other applications such as ambipolar transistor and optoelectronics, paving the way for new device architectures and high-performance devices.

**Introduction**

The advent of van der Waals (vdW) heterostructures has revolutionized materials physics and electronic and optoelectronic device technologies.[1~4] While vdW heterostructures are typically constructed from layered materials held together by vdW forces, integration of conventional materials in such structures is also desirable but carries unique challenges. Conventional metals are often used as electrical contacts and interconnects, however conventional metal deposition techniques can damage the vdW interface causing undesirable disorder.[5,6] Evaporated metal atoms with high kinetic energy bombard the contact areas during the evaporation, which has been shown to induce significant amounts of damage and defect states into vdW semiconductors, resulting in a strong Fermi level pinning effect[7,8] This in turn causes the Schottky barrier height (SBH) to become insensitive to the metal workfunction, resulting in low efficiency carrier injection and large contact resistance, which greatly hampers the observation of novel physics and the development of 2D semiconductor-based electronics. To address these challenges, considerable efforts have been devoted to minimize both Fermi pinning and contact resistance during the past decade. Forming vdW interfaces between three-dimensional metals (In[9], Pd[10]) and 2D semiconductors has been demonstrated via precise control of the electron-beam evaporation. Growth of buffer layers with low formation energy, such as semimetals (Bi, Sb, $ZrTe_2$)[11~13] on contact regions before metallization has been demonstrated as another effective strategy of *n*-type ohmic contacts for monolayer $MoS_2$ in terms of reducing the possible surface defect states, minimize SBH and even pushing the contact resistance down to quantum limitation (42 Ω·μm)[12]. On the other hand, surface transfer doping the TMDs ($MoS_2$ and $WSe_2$) beneath the electrodes by high-electron-affinity transition metal oxides (TMOs) has been reported as an excellent approach to form the *p*-type ohmic contacts down to the cryogenic regime because the TMOs degenerately dope the TMDs but creates minimal defects.[14,15]

An emerging strategy is the pick-and-place mechanical transfer of pre-deposited metal contacts, which has been demonstrated to maintain the low-energy vdW interaction between metal electrodes and 2D materials, rather than chemical bonding, resulting in an ideal interface and tunable SBH (reach the Schottky-Mott limit[16]). However, this method has been limited only to selected high-work function metals (Au, Pt, Pd) which can be peeled off and transferred with a high success rate. Many of the industry-preferred metals (such as Ti, Cr) show a strong adhesion to the substrates and cannot be picked up easily with the standard dry transfer technologies.[16] A few works have attempted to resolve this limitation.[17~19] Sacrificial layers (Se or propylene carbonate (PPC)) were formed on the 2D semiconductors before the high-energy metallization to protect the 2D semiconductors. Afterwards, the sacrificial layers were removed by an annealing process at elevated temperatures (Se @ ~350ºC, PPC @ ~250 ºC) and the top metal electrodes collapsed onto the 2D semiconductors.[18,19] However, any sacrificial-layer process adds additional material and the potential for impurities at the vdW interface which might alter its electronics properties. The additional processing is also undesirable and may be incompatible with other processes. Se deposition still requires a high energy process which may be harmful to the monolayer TMDs. The spin-coated PPC method only works with shadow masks because the PPC film is soluble to the common resist developers. Therefore, it is desirable to develop a simple, universal metal electrode fabrication technology which is universally compatible with vdW heterostructure fabrication.

Here we demonstrate a universal pick-and-place transfer technology for metal electrodes without the need for a sacrificial layer. Polished hydrogenated diamond substrates are utilized as reusable substrates for metal electrode fabrication by standard lithographic techniques. The low-energy diamond surface enables transfer of a wide variety of metals; we demonstrate transfer of eight different elemental metals with work function ranging from 4.22 to 5.65 eV. Hydrogenated diamond has already shown great potential in both electronics[20] and spintronics due to the intriguing *p*-type surface conducting nature. Furthermore, due to the low adhesion nature of hydrogenated diamond surface, prepatterned metals can be easily peeled off by Poly (Bisphenol A carbonate) (PC) or PPC stamps then aligned and laminated onto 2D semiconductors to form the 2D-material-based devices. We characterized the vdW interface between transferred electrodes and 2D semiconductors ($WSe_2$ and $MoS_2$) with high-resolution transmission electron microscopy (HRTEM) and observe atomically clean and damage-free interfaces. We demonstrate the utility of the technique by fabricating field-effect transistors from monolayer transition metal dichalcogenides (TMDs) with three diamond-transferred electrodes of different metals (Pd, Bi and Ti), showing low-SBH contacts in both *n*-type (Bi and Ti for $MoS_2$) and *p*-type (Pd for $WSe_2$) operation. We extend our diamond-transfer technology to few layer TMDs to demonstrate device applications such as ambipolar transistors, Schottky barrier diodes and photodetectors. Our work not only realized vdW contacts for 2D semiconductors with a broad range of metals, but also paves the way for the wafer-scaled transferred electrodes compatible with the commercialized nanofabrication process.

**Results and Discussion**

Prior to the metal contacts fabrication, diamond substrates were polished by scaif wheel (*Technical Diamond Polishing*) to minimize the surface roughness down to sub nanometer regime. Then the hydrogen termination was conducted in a *Seki 6500* diamond deposition reactor, consisting of a 2.4 GHz microwave plasma assisted chemical vapor deposition chamber. The pre-polished diamond samples were loaded onto a polycrystalline diamond coated molybdenum carrier and exposed to 85 Torr, 4500 W hydrogen plasma, with an $H_2$ flow rate of 450 sccm at 800 ºC. In order to produce a locally smooth surface (avoid the etching pits during the termination) and minimize contamination oxygen in the chamber, two different concentrations of $CH_4$ were added to the plasma, with an initial flow of 2.1 sccm $CH_4$ for 1 min during sample heating at the start of plasma exposure, followed by 4.1 sccm for another 1 min. The $CH_4$ flow was then turned off and the plasma was slowly extinguished by gradually the microwave power to 3200 W over a period of approximately 2 minutes, and then turning it off.. (Fig. 1. (a)) The roughness of hydrogenated diamond surface has been investigated by atomic force microscopy (AFM) before the metallization. As illustrated in Fig. 1. (b), the polished hydrogenated diamond has been characterized, and the root-mean-square roughness are around 0.32 nm, which suggests no etching pits formation during the hydrogen plasma process, resulting in an excellent smooth surface condition. Another two polished hydrogenated diamonds have been characterized as well and both of them show low surface roughness. (Fig. S1) The low roughness of hydrogenated diamond surface is critical for the following electrode fabrication and integration in atomically flat vdW heterostructures. As shown in Fig. S2 (a) and (b), electrode deposition and transfer from a different rough diamond with a high density of etch pits generated during the hydrogen termination results in visible roughness imprinted in the electrode surfaces after picking up from the rough diamond substrates. The rough electrodes showed greatly reduced adhesion as well as increased the non-uniformity between the contacts and 2D semiconductors after lamination (Fig. S2 (c) and (d)).

Then metal electrodes were fabricated on the polished hydrogenated diamond substrates by conventional photolithography technologies and e-beam evaporation. It should be noted that, after development of photoresist before metal deposition, diamond substrates were *in-situ* exposed to an argon plasma for two seconds to remove possible resist residues. After the lift-off process, the patterned metals on hydrogenated diamonds have been picked up by PC stamps (The PC stamps had been heated up to 150℃ to make the PC film completely touch the metal electrodes, then cooled down to room temperature to achieve the detachment). All steps were performed in a glove box to minimize the oxidization especially for low work function metals such as Ti and Bi. The pick-up process was smooth and gentle as shown in the video in Supplementary, which can greatly minimize the cracks or folding of metals. Lastly the picked-up metals were laminated onto the target 2D semiconductor heterostructures, as illustrated in Fig.1 (c).

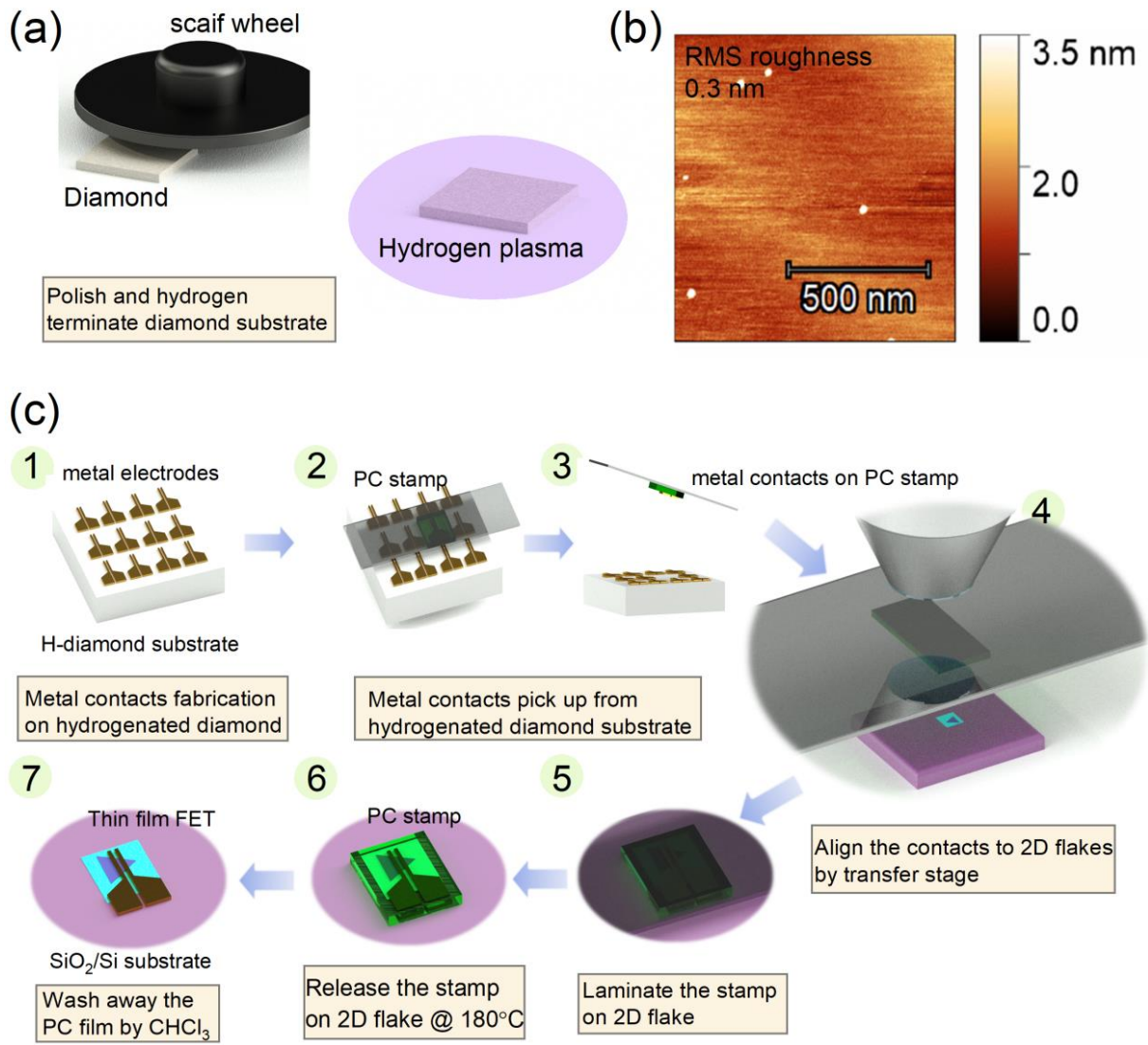

Figure 1. (a) Schematic illustration of diamond polishing and subsequent hydrogen termination. (b) AFM image for polished hydrogenated diamond surface. (c) Schematic illustration of hydrogenated diamond assisted metal electrode transfer for 2D semiconductor device fabrication.

In order to demonstrate this technique is effective for a broad range of metals, eight candidates (from high-work-function metal to low-work-function metal) have been investigated in this work. Fig. 2 shows optical micrographs of photolithography patterned metal films (Pt, Pd, Au, Ni, Cr, Ti, Al and Bi) on the diamond substrate, picked up by PC stamps, and transferred to target SiO$_2$/Si substrates, respectively. All metals can be peeled off from the hydrogenated diamond surface easily and keep their initial geometries without any cracks or folding after landing on SiO$_2$/Si substrates. Liu *et. al.* reported that metals with the strongest adhesion force (Ni, Ti and Cr) showed an extremely low releasing yield[16] from hexamethyldisilazane-treated silicon. Here, the diamond-assisted metal pick-up technology shows a very high releasing yield for these three reactive metals due to the low energy nature of hydrogenated diamond surface, demonstrating the universal nature of this technology. We have also shown that the diamond-assisted vdW electrode integration process is highly reproducible and scalable, making it suitable for large-scale 2D device fabrication. In Fig. S3 a ~ c, large-scale (3mm × 3mm) patterned electrodes (Cr 5nm/Au 20nm) was transferred from hydrogenated diamond to $SiO_2$/Si by PC method with nearly 100% yield. In addition, these electrodes array can also be aligned to the patterned films precisely as shown in Fig. S3 d and e, which paves the way for the wafer-scale vdW arbitrary-electrode integration for 2D semiconductor device fabrication.[23]

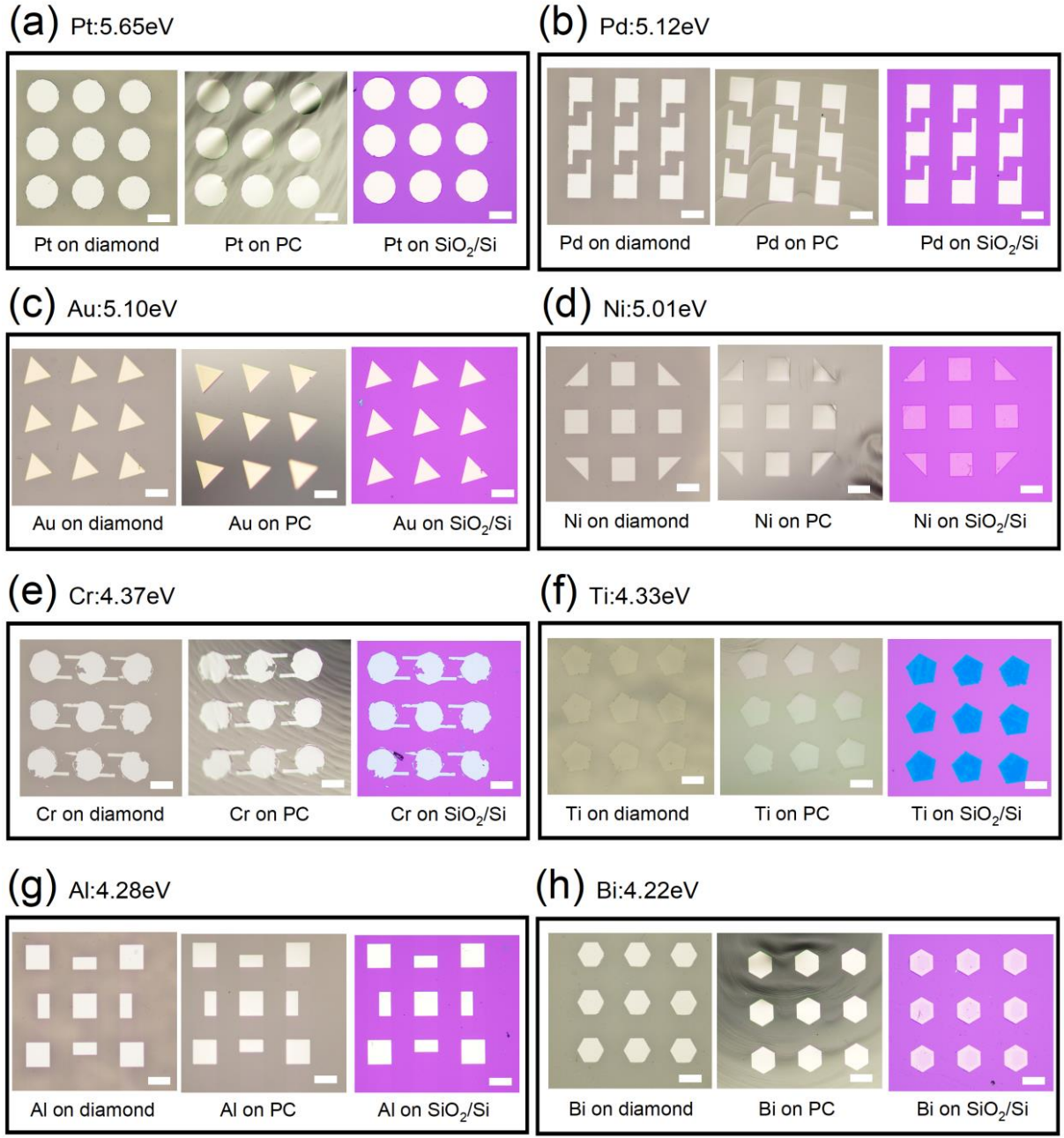

Figure 2. Micrographs of metal films (Pt, Pd, Au, Ni, Cr, Ti, Ai, Bi) in various shapes transferred from hydrogenated diamond substrates to SiO$_2$ (285nm)/Si substrates by dry transfer method. The white scale bars represent 50 µm.

We then utilized the diamond transfer technology (Fig. 1(c)) to integrate metal contacts on monolayer TMD/*h*-BN vdW heterostructures. We assessed the interface quality through cross-sectional TEM, energy-dispersive X-ray spectroscopy (EDS), while characterizing the device transport properties at both room temperature and 77K. According to Fig. 3 (a), the monolayer MoS$_2$ lattice structure is well retained without any damage, defects or contaminations (such as residual photoresist), which is consistent with the literatures.[16] Notably, the vdW gaps between metal/TMD and TMD/$h$BN were clearly discernible, as marked in Fig. 3 (b). In a few cross-section TEM images (Fig. S4), we find that a few localized bubbles of interstitial material are observed at the metal/monolayer TMDs interfaces. Eliminating such bubbles through techniques such as post annealing in Ar ambient or vacuum is the subject of future work. Consequently, employing vdW electrodes for contacting 2D semiconductors, especially for monolayer materials, presents a low-energy integration approach compared to conventional methods such as lithography and metallization. This approach significantly reduces the formation of defect states, strain, damage, and contamination associated with the lithography process.

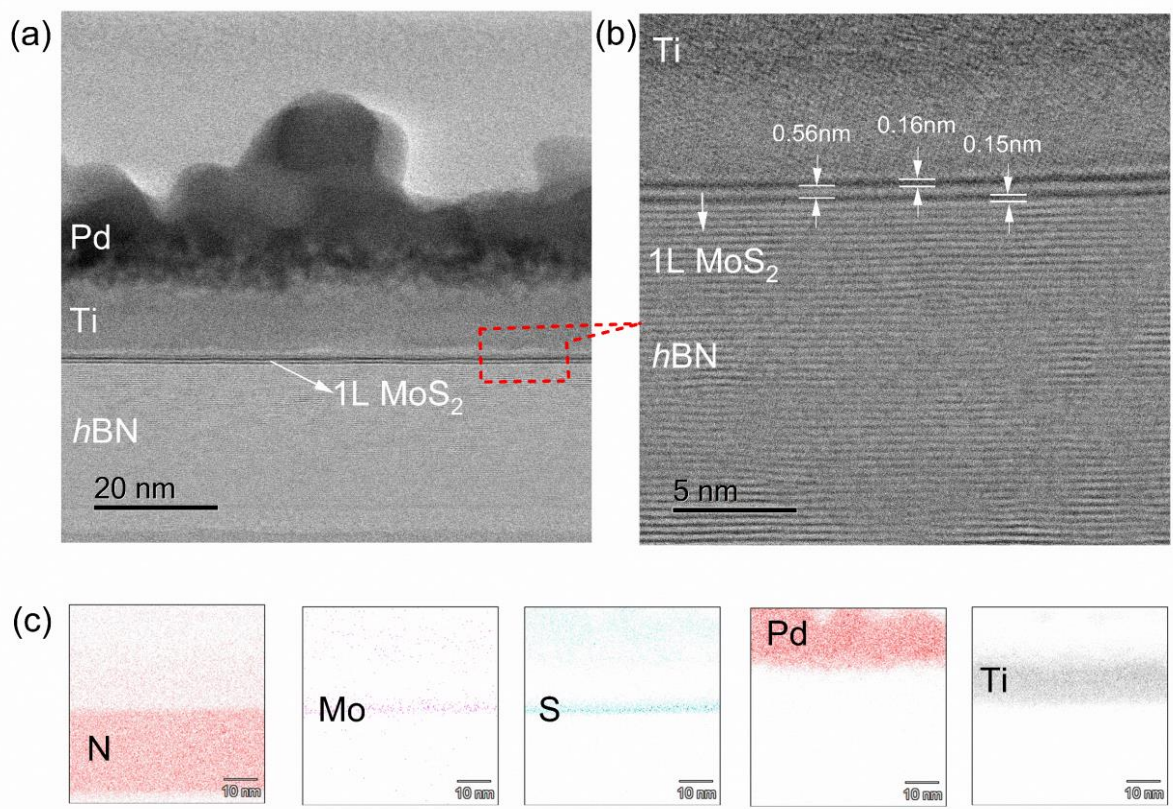

Figure 3. Characterization of the transferred electrodes (Pd/Ti) / monolayer TMD (MoS$_2$) / $h$BN heterostructures in terms of cross-section STEM using bright field imaging (a) and (b) and EDS elemental mapping (c).

In order to demonstrate the versatility of our technique, we fabricated both $n$-type and $p$-type back-gate 1L TMDs with three sets of transferred metal electrodes: (I) high-work-function metal Pd contact to 1L WSe$_2$ (optic image in Fig. 4a); (II) Semi-metal Bi contact to 1L MoS$_2$ (optic image in Fig. 4e); (III) Low-work-function metal Ti contact to 1L MoS$_2$ (optic image in Fig. 4i). Firstly, the Pd/WSe$_2$ FET shows typical hole conduction behavior with a high on/off current ratio of ~$10^7$. The source-drain current, $I_{sd}$, with -65 V gate bias at 77K exceeded that at room temperature. Two-terminal field-effect mobilities, as depicted in the inset of Fig. 4b, increased from 27 cm$^2$/Vs to 57 cm$^2$/Vs as the temperature dropped to 77K, indicating the good ohmic behavior between metals and TMDs. Output curves with gate bias of -65 V at both 300K and 77K displayed the high linearity, further suggesting the ohmic behavior with low Schottky barrier heights. Regarding the 1L-MoS$_2$ *n*-type FET, we also observed the ohmic behavior between transferred Pd/Bi and 1L MoS$_2$ in terms of transfer curves (Fig. 4f), field effect mobilities (inset of Fig. 4f) and linear output curves (Fig. 4g) at both room temperature and 77K, which is consistent with the reported works[11]. However, the 1L MoS$_2$ FET with Pd/Ti electrodes shows typical Schottky contact behavior. The source-drain current experiences a reduction during the cooling down process (Fig. 4j), and mobilities decrease from 3 cm$^2$/Vs to 0.6 cm$^2$/Vs as shown in Fig. 4k. The *IV* curve at a gate bias of 60V transitioned from linear at room temperature to nonlinear at 77K, indicating the presence of a high energy barrier (Fig. 4i). The possible reason might be due to the oxidation of Ti during the transfer process even in glove box environment. Despite this, all these three devices show high linearity in output curves (Fig. S5 b, d and f), suggesting the similar SBH at each metal-semiconductor interface, whereas the evaporated contacts often exhibit asymmetric *IV* due to the damage of the crystal lattice during the high-energy metal deposition process (Fig. S5 a, c and e). Regarding the reproducibility, we fabricated another three similar devices and studied the transport properties as a function of temperature. Pd/1L WSe$_2$ and Pd/Bi MoS$_2$ show ohmic behavior down to 77K (Fig. S6 a ~ f) whereas Pd/Ti MoS$_2$ shows Schottky behavior at low temperature (Fig. S6 g ~ i), which demonstrate the reproducibility of this contact fabrication technology. From Fig. S7, the Arrhenius plots for these three devices were further extracted, allowing us to estimate the SBH for each metal/TMD interfaces. The detailed calculated SBH at different gate bias extracted from the Arrhenius plots were plotted in Fig. 4d, h and i, separately. Both Pd/WSe$_2$ and Bi/MoS$_2$ interfaces show very low SBH, approximately 42 meV and 50 meV, respectively. These low SBHs are consistent with the linear *IV* curve at low temperatures. The high SHB (185 meV) between Ti and MoS$_2$ explains the nonlinear *I-V* behavior at low temperature. Notably, we demonstrate that different work function transferred metal electrodes can form low SBH to different type of TMDs, suggesting the negligible Fermi pinning effect can be achieved by diamond-assisted electrode transfer technology.

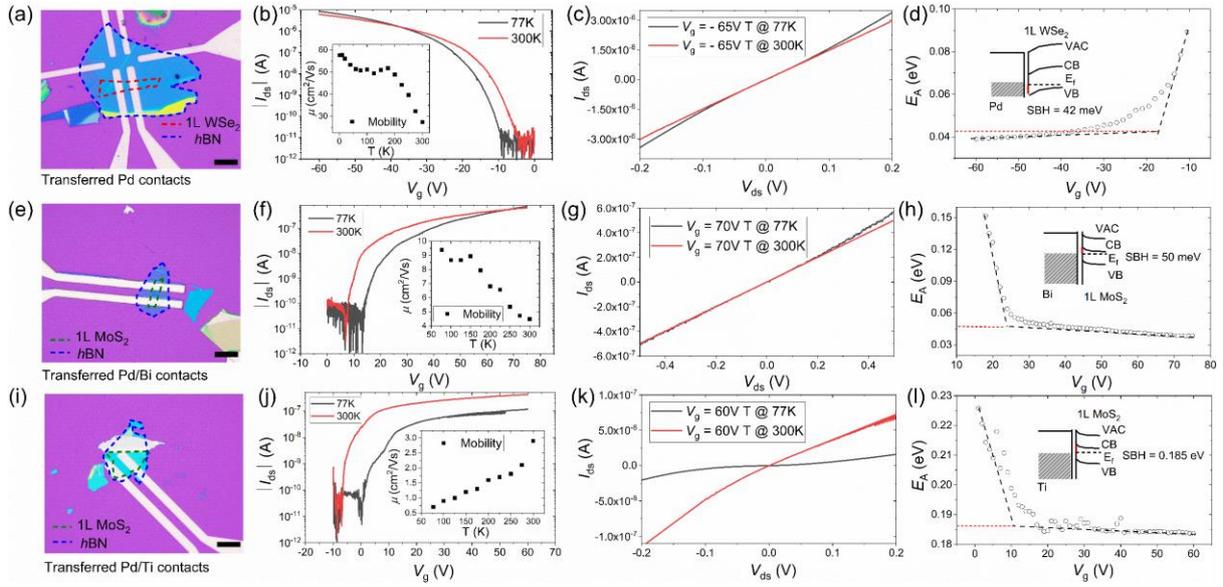

Figure 4. Transport characterization of transferred electrodes/monolayer TMDs FETs, including transfer curves, field-effect mobilities as a function of temperatures, output curves at different temperatures (77K and 300K) and Schottky barrier high analysis. (a) represents the optical image and (b) to (d) represent the data for 1L $WSe_2$ integrated with transferred Pd. (e) represents the optical image and (f) to (h) represent the data for 1L $MoS_2$ integrated with transferred Pd/Bi. (i) represents the optical image and (j) to (l) represent the data for 1L $MoS_2$ integrated with transferred Pd/Ti. The scale bar for each optical image represents 10 μm.

This straightforward diamond-assisted technique is not only confined to monolayer TMDs FETs, but also holds promise for other few-layer TMDs in diverse applications such as ambipolar transistors, Schottky barrier diodes and photodetectors. We firstly integrated Pd contacts to multilayer $WSe_2$ on $SiO_2$ (285nm)/Si substrate as depicted in the inset of Fig. 5a, and clearly demonstrated the symmetric ambipolar carrier transport according to the typical transfer curves (Fig. 5a) at room temperature. Fig. 5 b and c illustrate the linear output curves at room temperature, demonstrating the capability to modulate the conductance of $WSe_2$ channel from both negative (hole conduction) and positive (electron conduction) gate bias. For Schottky barrier diode application, we fabricated asymmetric contacts for a few-layer $WSe_2$ Schottky barrier diode (inset of Fig. 5d) through diamond-assisted electrode transfer technology. A pair of asymmetric electrodes, comprising high-work-function metal Au and low-work-function semimetal Bi, was transferred from diamond to a few-layer $WSe_2$ flake on $SiO_2$ (285 nm)/Si substrate. Fig. 5d shows the transfer curves of this device under different source-drain voltages. Here, the Au electrode was connected to the electrical ground, with source-drain voltage applied to Pd/Bi. The device displayed hole transport at negative gate bias, with the electron transport significantly suppressed due to the large Schottky barrier between Au and $WSe_2$. The output curves, as shown in Fig. 5e, exhibit pronounced rectification behavior depending on the gate bias ranging from -10 V to -60 V, with a high rectification ratio of about $10^2$ at a low source-drain voltage of $V_{ds} = \pm 0.5$V, suggesting a significant presence of contact resistance due to the reverse-biased metal-semiconductor junction.[24,25] The Schottky barrier diode with transferred electrodes demonstrates that the Fermi-level pining effect can be minimized with vdW contacts to 2D materials, allowing significant control of the SBH via choice of metal. As discussed in the transport studies of monolayer TMD devices, evaporated metal electrodes often suffer from unintentional variation of SBH from electrode to electrode. Here, we demonstrate that the diamond-transfer method results in reliably uniform SBH. The measurements details are discussed in the Methods and Experiments section. Three bilayer $WS_2$ photodetectors were fabricated with evaporated Pd, and one was integrated with transferred Pd. According to Fig. 5f, the open circuit voltage, $V_{oc}$, is negligible for the device with transferred contacts, however, the evaporated contacts devices exhibited varying $V_{oc}$, ranging from 0.05 V to 0.3 V. The significant shift in $V_{oc}$ suggests the uncontrollable asymmetric back-to-back SBH for evaporated electrodes, while the negligible $V_{oc}$ of transferred electrodes indicates the symmetric SBH. To quantify the photodetection efficiency, the photoresponsivity ($R$) was extracted for each device. The $R$ of transferred contact device is approximately 5 times larger than that of devices with evaporated contacts, as shown in Fig. 5g, indicating the superior photodetection efficiency of the device with transferred contacts.

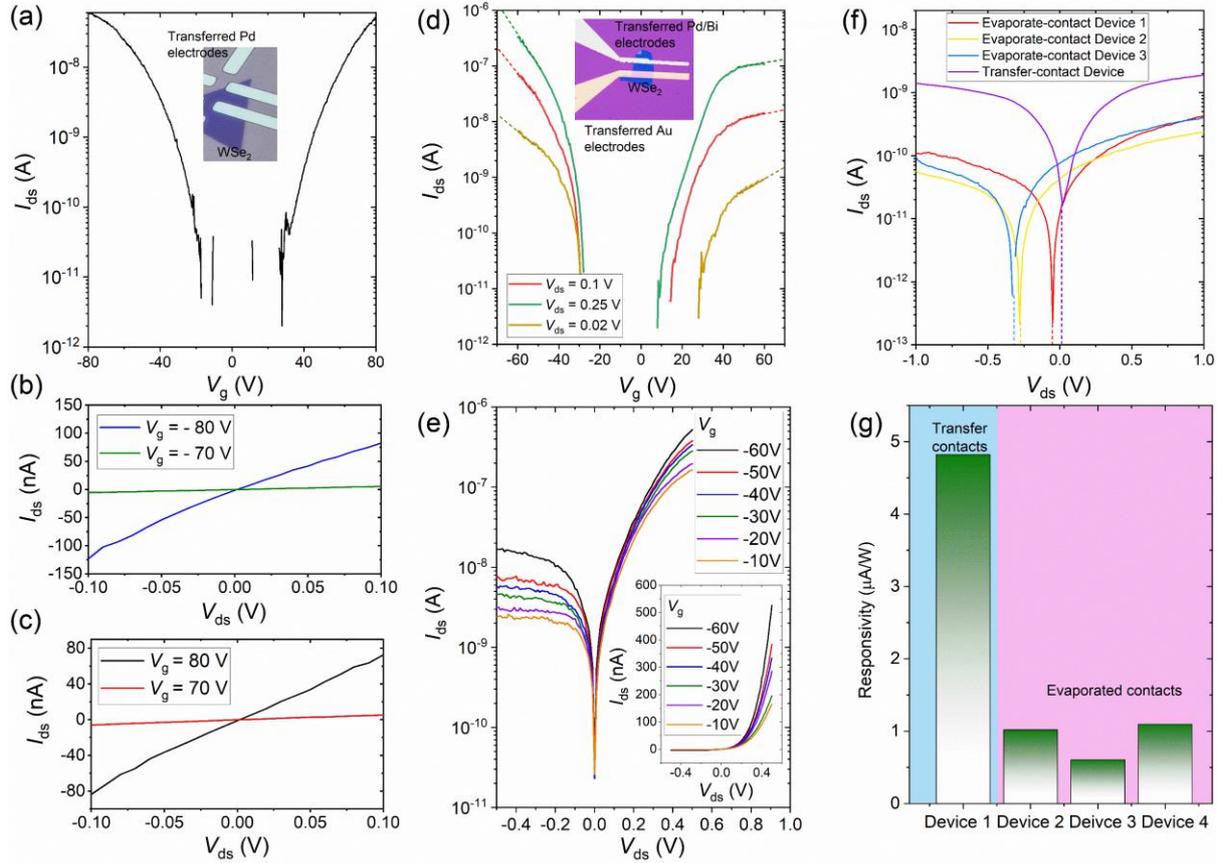

Figure 5. Transferred electrodes for few-layer TMDs applied to create an ambipolar FET, a Schottky barrier diode and a photodetector. (a) to (c) show the transfer characteristics (a) and output characteristics (b, c) in the *n*-type (b) and *p*-type (c) regimes of an ambipolar few-layer $WSe_2$ FETs with transferred Pd contacts. (d) and (e) show the transfer (d) and output characteristics of a Schottky barrier diode created by contacting few-layer $WS_2$ with asymmetric contacts (Pd/Bi and Au). (f, g) show the output characteristics (f) and photodetector responsivity (g) of few layer $WS_2$ photodetector devices with transferred and evaporated electrodes as indicated in the legends.

**Conclusion**

In summary, we have demonstrated a universal metal electrode transfer technology for contacting 2D semiconductors without the need for any buffer layers. By employing hydrogenated diamond surfaces, we successfully transferred eight different patterned metals, including high-work-function metals (Pt, Pd, Au, Ni), low-work-function metals (Cr, Al, Ti), and semimetals (Bi). The ease of peeling off from the hydrogenated diamond surface, thanks to its low adhesion nature, preserved their initial geometries without cracks or folding.

The formation of atomically smooth van der Waals (vdW) interfaces on 2D semiconductors was directly observed by TEM. The clean and smooth vdW interfaces effectively minimize the Fermi pinning effect and allow significant control of the Schottky barrier height through choice of metal.

We expect that our technology should be compatible with a wide range of other device applications, paving the way for new device architectures and high-performance devices. Additionally, it may provide an alternative fabrication process for air-sensitive materials such as $WTe_2$ and $MnBi_2Te_4$, opening avenues for exploring the new physics. Furthermore, this technology holds significant promise for wafer-scale fabrication with the advancement of large-area diamond-based wafers, offering exciting opportunities for large-scale production in the future.

**Methods and Experiments**

**Fabrication of vdW electrodes/monolayer TMDs/*h*BN/SiO$_2$/Si back gate transistors**

For TMDs device fabrication, high-work-function metal (Pd), low-work-function metal (Ti) and semimetal (Bi) were chosen to characterize the device performance. After fabricating the metal electrodes on hydrogenated surface (Fig.1 c, steps 1 and 2) by standard photolithography, *h*BN flakes were mechanically exfoliated onto a silicon substrate with 285 nm $SiO_2$ as dielectric layer by the blue tape (P/N: 1009R-1.0) technique. Monolayer TMD flakes ($WSe_2$ and $MoS_2$) were mechanically exfoliated onto polydimethylsiloxane (PDMS) and their monolayer nature was verified by both optical contrast and photoluminescence spectroscopy. After annealing the *h*BN dielectric in furnace (500 sccm Ar and 50 sccm $O_2$ at 500 ⁰C) for 3 hours, the targeted monolayer TMDs were transferred on the precleaned *h*BN by dry-transfer technique. Then the TMD/*h*BN heterostructures were annealed (950 sccm Ar and 50 sccm $H_2$ at 300 ⁰C) in furnace for 1 hour to remove most of the PDMS residue. Afterwards, the selected metal contacts were picked up by using a PC stamp as shown in step 3 of Fig. 1c and transferred onto the precleaned TMD/*h*BN heterostructures (Steps 4, 5 and 6 shown in Fig. 1c). In this work, Pt, Pd, Au, Ni, Cr, Ti, Al were deposited by e-beam evaporator (~ $10^{-6}$ mbar). Bi was thermally evaporated by using a molecular beam epitaxy system (~ $10^{-9}$ mbar). Lastly the PC film was removed by soaking in $CHCl_3$ at 60 ⁰C for 15 minutes to form the TMD FETs as shown in step 7 of Fig. 1c to expose the transferred electrodes for probing or wire bonding.

**FIB STEM/TEM and EDS measurements**

To ensure the preparation of a high-quality focused ion beam (FIB) lamella from vdW electrodes/TMD/$h$BN, an FEI HELIOS G3 microscope was employed for FIB lamella preparation. FIB lamella preparation followed the procedure developed for ion-beam-sensitive crystals reported in previous work[26]. A JEOL ARM 200F STEM was employed to acquire high resolution high angle annular dark field (HAADF) and bright field (BF) STEM images. EDS mapping was performed to confirm the composition of each layer in vdW electrodes/TMD/$h$BN.

**Electrical transport measurements**

Transport beahvior of the TMD transistors and Schottky barrier diode with vdW electrodes were measured in the temperature range of 300k to 77k using a TeslatronPT Oxford Cryostat. Two Keithley 2400 source meters were employed to apply the source-drain voltage ($V_{ds}$), gate voltage ($V_{gs}$) and monitor the source-drain current ($I_{ds}$), gate leakage current.

**Optoelectrical measurements**

The measurements were performed using a confocal microscope system (WITec alpha 300R) with a 50 × objective lens (NA = 0.9) in ambient condition. A 532 nm laser was fiber coupled through a fiber bench with optical chopper to adjust the power intensity and give the final spot size of 1 µm diameter. The samples were illuminated from the top side on a piezo-crystal-controlled scanning stage. One source meter was employed to apply the source-drain voltage and monitor the photocurrent.

**Acknowledges**

K. Xing, T.-H.-Y. Vu, M. Edmonds and M. Fuhrer acknowledge support from ARC grant DP200101345. D. McEwen, W. Zhao, M. S. Fuhrer acknowledge support from the ARC Centre of Excellence in Future Low-Energy Electronics Technologies (FLEET; CE170100039). M. Edmonds acknowledges the support from ARC grant FT220100290. D.-C. Q. acknowledges the support of the Australian Research Council (Grant No. DP230101904). D.-C. Q. acknowledge continued support from the Queensland University of Technology (QUT) through the Centre for Materials Science. A. Stacey acknowledges support of the Australian Research Council. (Grant No. LP190100528. )K.Watanabe and T.Taniguchi acknowledge support from the JSPS KAKENHI (Grant Numbers 21H05233 and 23H02052) and World

Premier International Research Center Initiative (WPI), MEXT, Japan. Q. Ou acknowledges support from the Science and Technology Development Fund, Macau SAR (No. 0065/2023/AFJ, No. 0116/2022/A3, No. 0009/2022/AGJ). The microscopy was enabled by the Electron Microscopy Centre at the University of Wollongong and used the FEI Helios G3 CX funded by the ARC LIEF grant (No. LE160100063), JEOL JEM-ARM200F funded by the ARC LIEF grant (No. LE120100104). This work was performed in part at the Melbourne Centre for Nanofabrication (MCN) in the Victorian Node of the Australian National Fabrication Facility (ANFF) with M. Edmonds ANFF-VIC Technology Fellowship.

**Author contributions**

K. Xing, D.-C. Qi and M.S. Fuhrer conceived this research. K. Xing designed the experiments. K. Xing, D. McEwen and A. Stacey lead the samples fabrication with the assistance of T.-H.-Y. Vu. K. Xing and W. Zhao lead the electrical characterization. A. Bake and D. Cortie performed the cross-section HRTEM and EDS measurements. M.T. Edmonds helped with the Bi growth. J. Hone provided the TMD ($WSe_2$ and $MoS_2$) crystals. K. Watanabe and T. Taniguchi provided the $h$BN crystals. K. Xing, J. Liu and Q. Ou lead the photodetector measurements. The manuscript was written through contributions of all authors. All authors have given approval to the final version of this manuscript.

**Additional information**

Supplementary information: The online version contains supplementary information available at